\documentstyle[12pt]{article}
\def\bar{\overline}
\def\ZZ{\mbox{\rm Z}\hskip-4pt \mbox{\rm Z}}
\def\RR{\mbox{\rm I}\hskip-2pt \mbox{\rm R}}

\def\id{\mbox{\rm 1}\hskip-3pt \mbox{\rm l}}

\newcommand{\bg}{\begin{equation}}
\newcommand{\eg}{\end{equation}}
\def\su21{$SU(2\, \vert \, 1)$}

\begin{document}
\begin{center}
{\Large ALGEBRAIC CONNECTIONS ON\\[6pt]
PARALLEL UNIVERSES$^{\ast )}$}
\end{center}
\vspace{2cm}
\begin{center}
R. Coquereaux$^1$\\
{\it MPCE, Macquarie University, New South Wales, Australia}\\[1cm]
R. H\H au\ss ling, F. Scheck\\
{\it Institut f\H ur Physik\\Johannes Gutenberg-Universit\H at\\
D-55099 Mainz (Germany)}
\end{center}
\vspace{4cm}

\begin{abstract}
For any manifold $M$, we introduce a $\ZZ $-graded differential algebra
$\Xi$, which, in particular, is a bi-module over the associative algebra
$C(M\cup M)$. We then introduce the corresponding covariant differentials
and show how this construction can be interpreted in terms of Yang-Mills
and Higgs fields. This is a particular example of noncommutative geometry.
It differs from the prescription of Connes in the following way: The
definition of $\Xi$ does not rely on a given Dirac-Yukawa operator acting
on a space of spinors.
\end{abstract}

\vfill
{\small $^{\ast)}$ Work supported in part by PROCOPE project
Mainz University and CPT Marseille-Luminy\\
CPT 93/PE 2947 \\
$^{1}$ On leave of absence from Centre de Physique Th\'eorique, CNRS Luminy,
Case 907, F-13288 Marseille}

\newpage
{\bf 1. Introduction}\\[6pt]
In physics, fundamental interactions are described in terms of covariant
differentials, or connections. These covariant differentials appear
because we want physics to
be independent of the choice that we make, at every point of space-time
$M$, of a particular ``frame''. This ``frame'' can be a set of four
independent vectors belonging to the tangent space at $x\in M$ (we then
describe gravity) but it can also be a set of three independent complex
vectors, smoothly dependent on $x\in M$ (we then describe
chromodynamics). For electroweak interactions the situation is similar.\par
In all cases, the freedom of choosing different frames (``external'' or
``internal'') at different space-time points leads to gauge fields
(connection one-forms) appearing in the covariant derivatives and to
different kinds of ``forces'', described by the corresponding curvatures
(assuming that our ``external'' and ``internal'' space-time is curved!).
In mathematical language, and at the classical level, the fundamental
interactions of physics are described by connections in
appropriate principal (or vector) bundles.\par
There is however an important aspect of physics that is not described by
the above mathematical structure: There are scales (or masses) in physics.
In all classical treatments, gauge fields and matter fields have a nice
geometrical interpretation but mass-related phenomena do not. In particle
physics, the Higgs mechanism is taken as ultimately responsible for
mass generation, but the Higgs field along with its self-coupling did
not appear as very appealing -- from the aesthetical point of view -- when
it was proposed more then twenty years ago.\par
Recently, several constructions of the fundamental interactions using
``noncommutative geometry'' were proposed
\cite{Co1,Co2,CEV,CES,CFF,DV}. All these constructions share the following
common features:\par
\begin{itemize}
\item The concepts of covariant derivative and connection are extended
from the realm of manifolds (or commutative algebras) to arbitrary
associative (not necessarily commutative) algebras. At the same time, the
usual algebra of differential forms -- that plays a key role in the usual
construction -- is replaced by some more general differential algebra.
\item The Higgs field appears as part of the generalized connection, and
its coupling to the Yang-Mills fields and the Higgs potential
itself appear naturally in the expression of the generalized Yang-Mills
functional (the trace of the squared generalized curvature).
\item Mass generation for fermions, via Yukawa couplings to the Higgs, is
now incorporated in a generalized Dirac operator that we shall call the
Dirac-Yukawa operator.
\end{itemize}
The different constructions that were proposed are described in
the literature and we do not review them here.
The purpose of the present paper is to explain how the specific
construction that was presented in
\cite{CEV} and continued in \cite{CES} fits into the general framework of
noncommutative geometry. Indeed, \cite{CEV} was written on purpose in a way
that involves only elementary mathematical tools in order for
the construction to be understood without having to rely on a number of
rather nontrivial mathematical concepts.\par
However, the mathematically oriented reader might wish to know if the
construction presented in \cite{CEV} is indeed a particular case of a
``noncommutative connection'' or if it is something different; he will want
to obtain answers to the following questions:\par
\begin{itemize}
\item What is the underlying associative algebra?
\item What is the underlying differential algebra?
\item What is the module?
\end{itemize}
We shall answer these questions here.\par
The present paper is not conceived as a sequel to \cite{CEV} (or \cite{CES}).
It is selfcontained, to a large extent, and, therefore, it is of
independent interest.\par
Furthermore, for any manifold, we shall construct a new $\ZZ $-graded
differential algebra by using matrix-valued differential forms (cf. sec. 3).
This algebra is neither commutative nor
even graded-commutative (and therefore does not  coincide with the usual
algebra of differential forms); it underlies, albeit not explicitly, the
construction made in \cite{CEV} and is very similar to the algebra
$\Omega_D (A)$ introduced by Connes in \cite{Co2}.\\[6pt]

{\bf 2. Connections in noncommutative geometry}\\[6pt]
In order to define covariant differentials in noncommutative geometry,
one needs three ingredients \cite{Co3}:\par
\begin{enumerate}
\item an  algebra $A$, not necessarily commutative;
\item a $\ZZ $-graded  differential algebra
$\left( \Xi = \oplus^\infty_{p=0}\Xi^p,d\right)$ such that $\Xi^0 = A$.
Notice that $\Xi$  is, therefore, an $A$-bimodule.
\item a right-module ${\cal S}$ over $A$. (This could actually be a
left-module but our choice makes notations easier.)
\end{enumerate}
In conventional gauge theories, i.e. in the framework of commutative
geometry, the ingredients
are as follows:\par
\begin{enumerate}
\item $A$ is the (commutative) algebra $C(M)$ of real or complex  valued
functions on a smooth manifold $M$, with $M$ representing, for example,
space-time,
\item $\Xi$ is the algebra of differential forms $\Lambda (M)$  with its
wedge product, and
\item ${\cal S}$ is the space of sections of a vector bundle (for instance
the space of vector fields, scalar fields, tensor fields, spinor fields,
quark fields).
\end{enumerate}
The covariant differential $\nabla $ maps ${\cal S}$ into ${\cal S}
\otimes_A \Xi^1 $, and more generally ${\cal S} \otimes_A \Xi^p $ to
${\cal S} \otimes_A \Xi^{p+1} $. For instance, it maps vector fields
to one-form valued vector fields. $\nabla $ should be such that\par
\begin{equation}
\nabla (vf) = (\nabla v) f + v\otimes df \qquad , v\in {\cal S}, f\in A
\end{equation}
and more generally\par
\begin{equation}
\nabla (v\otimes \lambda ) = (\nabla v)\lambda
                             + (-1)^p v\otimes d\lambda \qquad ,
                             v\in {\cal S}, \lambda \in \Xi^p .
\end{equation}
In usual differential geometry we can define not only covariant
differentials but also covariant derivatives by evaluating the covariant
differential $\nabla v$ along a vector field $e_{\mu}$:
$\nabla_{e_{\mu}}v$. Indeed, in the
usual case, the space of one-forms is dual to the space of vector fields
and vector fields themselves are derivations on the commutative algebra
$C^\infty (M)$. However, an arbitrary algebra may not have derivations at all
(for instance the complex numbers); of course one could argue that
 ``interesting'' algebras have non trivial derivations
(for instance the inner derivations). The problem is that, in any case,
derivations on the algebra $A$, even if they exist,  will not form
a module over $A$ (unless we are in the commutative or
 graded-commutative case).
Finally there is no reason why these derivations should be related to
one-forms (elements of $\Xi^1 $).
To summarize: we have covariant differentials but no covariant derivatives
(unless we introduce a space dual (over $A$) to $\Xi^1$ and evaluate $\nabla v$
on its elements.)\par
It should be stressed that, for a given algebra $A$ (commutative or not),
every choice of a $\ZZ$-graded differential algebra $\Xi$ defines a
differential calculus.\par
Many nice properties of covariant differentials can be generalized to the
noncommutative case \cite{Co3}. In particular the square $\nabla^2$ of the
covariant
differential is an $A$-linear object: the curvature. Let us suppose that
${\cal S}$ can be written as ${\cal S} = pA^n$ where $p$ is a projector
($p^2 = p$). This is a mild assumption since all the modules obtained as
spaces of sections of vector bundles are of this form (projective finite
modules).  The case $p = Identity$  means that the module (the vector
bundle) is trivial. A covariant differential can be written as
\begin{equation}
\nabla = pd + {\cal A}
\end{equation}
where ${\cal A} \in \Xi^1 \otimes_{\cal A} End{\cal S}$
 and should be such that $p{\cal A}p = {\cal A}$.
Then, if $X\in {\cal S}$, $\nabla X = pdX + {\cal A}X$, which, when written
in components, reads $(\nabla X)^i = p^i_k (dX)^k + {\cal A}^i_k X^k$.
It is easy to check that $\nabla$ is indeed a connection, viz.
\begin{eqnarray}
\nabla (Xf) & = & p d(Xf) + {\cal A} X f \nonumber\\
            & = & p (dX) f + {\cal A} X f + p X df \nonumber\\
            & = & \nabla (X) f + X df
\end{eqnarray}
where we used the fact that $pX = X$. Then we compute
\begin{eqnarray}
{\cal F} & = & \nabla^2 \\
\nabla^2 (X) & = & pd(pdX + {\cal A}X) + {\cal A}(pdX + {\cal A}X)
                   \nonumber\\
             & = & pd(pdX + {\cal A}X) + {\cal A}(dX + {\cal A}X),
                   \,\mbox{since } {\cal A}p = {\cal A} \nonumber\\
             & = & p(dp dX + pd^2 X + (d{\cal A})X - {\cal A}(dX)
                   + {\cal A}(dX)) + {\cal A}^2 X \nonumber \\
             & = & p(dp dX + (d{\cal A})X) + {\cal A}^2 X \nonumber\\
             & = & [pdpdp + pd{\cal A} + {\cal A}^2 ]X
\end{eqnarray}
where we have used the properties $dX = d(pX) = (dp)X + pdX$ and
$p^{2}(dp)p = pd(p^2) - pdp = 0$ implying $pdpdX = pdpdpX + p(dp)pX = pdpdpX$.
The conclusion is that the curvature is
\begin{equation}
{\cal F} = pdpdp + p d{\cal A} + {\cal A}^2
\end{equation}
and is linear.\par
Notice that the associative algebra $A$ could be $\ZZ_2$-graded; the whole
formalism of noncommutative geometry extends to the $\ZZ_2$-graded case and
this is discussed in \cite{Kas}.\\[6pt]

{\bf 3. A simple example of generalized connections}\\[6pt]
We now construct a very simple example that fits into the general
framework but is slightly more general than the situation of pure
Yang-Mills theory.\par
For the associative algebra $A$ we take $A = C(M)\oplus C(M)$ and we
represent elements of $A$ as $2\times 2$ diagonal matrices with elements
$f(x)$ und $g(x)$, two numerical functions over the $n$-dimensional space
(or space-time $M$). Notice that $A$ is still commutative but does not
coincide with the space of functions over a connected manifold: Indeed $A$
is the space of functions over the non-connected disjoint union
$M\cup M$. Intuitively we can think of it as two parallel universes where
left and right ``movers'' live (or where fermionic particles and
antiparticles live \cite{Bond}).\par
For ${\cal S}$ we take ${\cal S} = A$ itself,
represented as the column vectors with two elements $f(x)$
and $g(x)$. An alternative choice, motivated by physical considerations,
would be to take a space of vectors with two components $\Psi_L$ and
$\Psi_R$, the first being a left-handed spinor field and the second
being a right-handed spinor field. (Of course, here we have to assume
that $M$ is even dimensional and admits a spin structure.) \par
To fully set the stage we have to specify the $\ZZ$-graded differential
algebra $\Xi$. This is where our construction differs from
\cite{Co1}. Let us just define $\Xi$ as a vector space. (To simplify the
description we assume here that $M$ is $4$-dimensional, but generalizing
the construction for any dimension $n$, $n$ even or odd, is obvious.)\par
\begin{itemize}
\item $\Xi^0$ is the space of $2\times 2$ matrices with zero-forms
(functions) on the diagonal.
\item $\Xi^1$ is the space of $2\times 2$ matrices with one-forms
on the diagonal and zero-forms on the antidiagonal.
\item $\Xi^2$ is the space of $2\times 2$ matrices with zero-forms and
two-forms on the diagonal and one-forms on the antidiagonal.
\item $\Xi^3$ is the space of $2\times 2$ matrices with one-forms and
three-forms on the diagonal and zero-forms and two-forms on the antidiagonal.
\item $\Xi^4$ is the space of $2\times 2$ matrices with zero-forms and
two-forms and four-forms on the diagonal and one-forms and three-forms
on the antidiagonal.
\item $\Xi^5$ is the space of
$2\times 2$ matrices with one-forms and three-forms on the diagonal and
zero-forms and two-forms and four-forms on the antidiagonal.
\end{itemize}
Note that this sequence of spaces does not stop at $\Xi^4$, even though
$\mbox{dim }M$ is 4. For the remainder, we take
\begin{eqnarray*}
\Xi^6 \cong \Xi^8 \cong \Xi^{10} & \cong & \ldots \mbox{ as isomorphic
copies of } \Xi^4 \mbox{ and}\\
\Xi^7 \cong \Xi^9 \cong \Xi^{11} & \cong & \ldots \mbox{ as isomorphic
copies of } \Xi^5 .
\end{eqnarray*}
We then define $\Xi$ itself as the direct sum
$\Xi = \oplus^\infty_{p=0} \Xi^p$.\par
It is to be noted that we define $\Xi$ as a {\it direct sum} of vector
spaces. This implies, in particular, that $\Xi^p \cap \Xi^q = \emptyset$
whenever $p\neq q$, and, for instance, that $\Xi^2$ is not
included in $\Xi^4$! This direct sum construction is exactly analogous to
writing
$\RR^2 = \RR \oplus \RR$ for the two dimensional plane. Clearly, the
first copy of $\RR$ is
isomorphic to the second one but the $\partial_x$- and $\partial_y$-axes
are different. It is probably not necessary to elaborate more on such a
elementary remark but the fact itself is essential.\par
For a smooth manifold $M$ of dimension $n$ we call $\Lambda(M)$ the algebra of
differential forms and define $\Xi^{2p}$ as the
space of $2\times 2$ matrices such that diagonal elements belong to
$\oplus^{2p}_{j=0,(even)} \Lambda^j (M)$ and such that antidiagonal
elements belong to $\oplus^{2p-1}_{j=1,(odd)} \Lambda^j (M)$;
similarly we
define $\Xi^{2p+1}$ as the space of $2\times 2$ matrices such that diagonal
elements belong to $\oplus^{2p+1}_{j=1,(odd)} \Lambda^j (M)$
and such that
antidiagonal elements belong to $\oplus^{2p}_{j=0,(even)}
\Lambda^j (M)$.\par
As a vector space we have in all cases
\begin{equation}
\Xi = \oplus^\infty_{q=0} \Xi^q
\end{equation}
with
\begin{equation}
\Xi^{2p} = \left[ \oplus^{2p}_{j=0,(even)}
                               \Lambda^j (M) \right] \oplus
           \left[ \oplus^{2p}_{j=0,(even)}
                               \Lambda^j (M) \right] \oplus
           \left[ \oplus^{2p-1}_{j=1,(odd)}
                               \Lambda^j (M) \right] \oplus
           \left[ \oplus^{2p-1}_{j=1,(odd)}
                               \Lambda^j (M) \right]
\end{equation}
and a similar expression for $\Xi^{2p+1}$. (Notice that, $M$ being finite
dimensional, we always get a periodicity modulo 2 for $\Xi^q$ when $q$ is
big enough.)\par
To make $\Xi$ an algebra, we define the following associative, graded
product $\odot$:
\begin{equation}
(M\otimes f) \odot (N\otimes g) = (-1)^{\partial N \partial f}
                                  MN\otimes f\wedge g
\end{equation}
where $\partial f$ denotes the $\ZZ_2$-grading of the form $f$ (even or odd)
 and $\partial N$ denotes the $\ZZ_2$-grading of the $2 \times 2$ matrix $N$
 (diagonal or antidiagonal).
For the case of $2\times 2$ matrices this reads
\begin{equation}
\left( \begin{array}{cc}
         A & C \\ D & B
       \end{array} \right) \odot \left( \begin{array}{cc}
                                          A' & C' \\ D' & B'
                                        \end{array} \right) =
\left( \begin{array}{cc}
         A\wedge A' + (-1)^{\partial C} C\wedge D' &
         C\wedge B' + (-1)^{\partial A} A\wedge C' \\
         D\wedge A' + (-1)^{\partial B} B\wedge D' &
         B\wedge B' + (-1)^{\partial D} D\wedge C'
       \end{array} \right)
\end{equation}
It should be understood that the product of an element of $\Xi^p$ and an
element of $\Xi^q$ belongs to $\Xi^{p+q}$.
It is easy to check that the product $\odot$ makes $\Xi$
a $\ZZ$-graded algebra.\\
Finally we want to define a differential $d$ on $\Xi$. (Actually it will be
a graded differential.) With
\begin{equation}
X = \left( \begin{array}{cc}
             A & C \\ D & B
           \end{array} \right) \in \Xi^p
\end{equation}
we define
\begin{eqnarray}
\delta_1 X & = & \left( \begin{array}{cc}
                            dA & -dC \\ -dD & dB
                          \end{array} \right) \in \Xi^{p+1} \mbox{ and} \\
\delta_2 X & = & i \left( \begin{array}{cc}
                              C+D & -(A-B) \\ (A-B) & C+D
                            \end{array} \right) \in \Xi^{p+1} \, ,
\end{eqnarray}
with $d$ denoting the usual exterior derivative.\\
It is easy to check that $\delta_1^{ 2} = \delta_2^{ 2} =
\delta_1 \delta_2 + \delta_2 \delta_1 = 0$. We therefore define
\begin{equation}
dX = \delta_1 X + \delta_2 X \in \Xi^{p+1}
\end{equation}
and verify that $d^2 X = 0$. Moreover,
\begin{eqnarray*}
\mbox{for } X\in \Xi^{2p}, Y\in \Xi^r \mbox{ we have }
d(X\odot Y) & = & dX\odot Y + X\odot dY , \\
\mbox{but for } X\in \Xi^{2p+1}, Y\in \Xi^r \mbox{ we have }
d(X\odot Y) & = & dX\odot Y - X\odot dY .
\end{eqnarray*}
Therefore $\Xi$ is a $\ZZ$-graded and $\ZZ_2$-graded differential algebra.
This completes the list of ingredients needed to construct an example of
gauge theory in noncommutative geometry.\par
As explained in section 2, an arbitrary element of $\Xi^1$ defines a
connection, moreover the module ${\cal S}$, in our case, is trivial, so
that the curvature is simply:
\begin{equation}
{\cal F} = d{\cal A} + {\cal A} \odot {\cal A} \in \Xi^2
\end{equation}
Writing
\begin{equation}
{\cal A} = \left( \begin{array}{cc}
                    L & i\phi \\ i\bar{\phi} & R
                  \end{array} \right) \in \Xi^1
\end{equation}
we obtain
\begin{eqnarray}
{\cal F} & = & \left( \begin{array}{cc}
                        {\cal F}_{11} & {\cal F}_{12} \\
                        {\cal F}_{21} & {\cal F}_{22}
                      \end{array} \right) \nonumber \\
\mbox{with } {\cal F}_{11} & = & dL - [(\phi + \bar{\phi})
                                       + \phi \bar{\phi}] \nonumber \\
{\cal F}_{12} & = & -i [d\phi + L\phi - \phi R + (L-R)] \nonumber \\
{\cal F}_{21} & = & -i [d\bar{\phi} - \bar{\phi} L + R\bar{\phi}
                        - (L-R)] \nonumber \\
{\cal F}_{22} & = & dR - [(\phi + \bar{\phi}) + \phi \bar{\phi}]
\end{eqnarray}
We now suppose that the underlying manifold is Riemannian
(or pseudo-Riemannian); therefore we have a metric $g = (g_{\mu \nu})$,
i.e. a scalar product on the tangent spaces. Assuming that it is not
degenerate, we can extend this scalar product to one-forms and more
generally to the whole algebra of differential forms. (The extension is
not unique in the sense that we could introduce arbitrary
positive constants, or rescaling factors, in the expression of $g$ for
$p$-forms and $q$-forms.)\par
Writing
\begin{equation}
{\cal L} = |{\cal F}_{11}|^2 + |{\cal F}_{12}|^2 +
           |{\cal F}_{21}|^2 + |{\cal F}_{22}|^2 ,
\end{equation}
we find (\cite{CEV}), up to an overall rescaling,:
\begin{eqnarray}
{\cal L} & = & - \frac{1}{4} (F_{\mu\nu}^L)^2 -
                 \frac{1}{4} (F_{\mu\nu}^R)^2
               + 2 \bar{D\phi} D\phi
               + 2 (\phi + \bar{\phi} + \phi \bar{\phi} )^2 \\
\mbox{with } D\phi & = & \nabla \phi + (L - R) \nonumber \\
                   & = & d\phi + L\phi - \phi R + (L - R)
\end{eqnarray}
and $F^{L}=dL$ and likewise for $R$.
It is then convenient to introduce new fields $\gamma$ and $Z$
\begin{eqnarray}
\gamma & = & \frac{1}{\sqrt{2}} (L+R) \nonumber \\
Z & = & \frac{1}{\sqrt{2}} (L-R) .
\end{eqnarray}
The above formalism describes therefore a $U(1)\times U(1)$ theory with
symmetry breaking. (The photon $\gamma$ stays massless but the $Z$
acquires a mass.) Notice that the potential
\begin{equation}
V(\phi ) = (\phi + \bar{\phi} + \phi \bar{\phi} )^2
\end{equation}
is already shifted to a point of absolute minima (see the discussion in
\cite{CEV,CHPS}).\par
At this point it is useful to point out the fact that the algebra
$\Omega_D (A)$ introduced in \cite{Co2} and the above algebra $\Xi$ are
very similar. The former is obtained from the universal differential
algebra $\Omega (A)$ by dividing out a differential ideal ${\cal J} =
{\cal J}_0 + \delta {\cal J}_0$ where ${\cal J}_0$ is the kernel of the map
\begin{equation}
\Omega (A) \ni a_0\delta a_1 \ldots \delta a_p \mapsto a_0[D,a_1]\ldots
[D,a_p] \in {\cal B}(H) ,
\end{equation}
$D$ being the Dirac-Yukawa operator and ${\cal B}(H)$ being the space of
bounded operators on the Hilbert space of spinors $H$. Actually ${\cal J}$
is not naturally graded and one has to grade it by taking its intersection
with the vector subspaces of $\Omega(A)$ associated with its $\ZZ$-grading.
 This division leads
to the consequence, in Connes' approach, that
the kinetic term $\bar{D\phi} D\phi$ is proportional to
tr$(MM^{\dagger})$ and
that the Higgs potential is
proportional to tr$[((MM^{\dagger})_{\perp})^2]$, where
\[\mbox{ }( MM^{\dagger} )_{\perp}=
\mbox{}(MM^{\dagger})-\frac{1}{n} \mbox{tr}MM^{\dagger}\, ,\]
where $M$ is a $n \times n$
fermionic mass matrix. Hence, this potential vanishes whenever
 $MM^{\dagger}$ is
proportional to the unit matrix.\par
The spaces $\Omega_D^0 (A)$, $\Omega_D^1 (A)$ and $\Omega_D^2 (A)$ for
$A = C(M) \oplus C(M)$ have been computed in \cite{Co2} and it happens that
they are respectively equal to $\Xi^0$, $\Xi^1$ and $\Xi^2$. This comparison
applies when $MM^{\dagger}\neq \id$, i.e., in Connes' model, when the
fermion masses are not degenerate. Note that in our construction no mass
matrix appears.\par
To avoid possible confusions, we should mention the fact that the notation
$\Omega_D(A)$ itself was introduced already in \cite{DVACAD} to denote another
differential algebra (namely, the homomorphic image of the universal
differential envelope of $A$ into the algebra of $A$-valued multilinear
forms on the space of derivations of $A$.)
\\[6pt]
{\bf 4. Conclusions}\\[6pt]
Connes' differential algebra $\Omega_D (A)$ depends explicitly on the
fermionic masses entering the Dirac-Yukawa operator $D$. Apart from this
difference, we have seen that the spaces that are relevant for physics,
i.e. the spaces of degree 0, 1, and 2, are the same in our construction.
In particular, the expression of ${\cal L}$ (the
bosonic Lagrangian) is esentially the same in both cases. The only
difference is the following: In our case, using $\Xi$, the coefficients in
front of the individual terms of the Lagrangian are arbitrary. They stem
from different normalizations of the scalar product in the space of $p$-
forms. In the case of \cite{Co2} (using $\Omega_D (A)$), these coefficients
are computed in terms of masses of fermions. It is not clear how
quantization could be made to respect, or modify in a predictable way, the
mass relations obtained using $\Omega_D (A)$. If one disregards these mass
relations, the two approaches become completely equivalent, as far as physics
and the standard model of electroweak interactions are concerned.
While the approach in \cite{CEV,CES} is more elementary and easily amenable
 to a
physical interpretation, the approach \cite{Co1,Co2} has a bigger character of
generality.\par
Also we would like to emphasize the following point: The above formulation
does not use explicitely the $\ZZ$-grading of the algebra $\Xi$;
everything can be done using only $2\times 2$ matrices, i.e. representing
the {\it whole\/} algebra $\Xi$ by $2\times 2$ matrices. In this process, the
$\ZZ$-grading is lost, only the $\ZZ_2$-grading remains (it is not a
representation of $\ZZ$-graded algebra). This is what was
done in \cite{CEV} because it is simpler. The drawback was that
the underlying $\ZZ$-graded algebra (which had to be there in order to put
things in a more conventional mathematical framework) was not obvious. The
main purpose of the present paper was to answer that question.\par
In the present paper we presented the example of $2\times 2$ matrices
because this makes the underlying structure clear. More generally, we
could use $2N\times 2N$ matrices and this would lead to a
$U(N)_L \times U(N)_R$ gauge theory. The structure group can
subsequently be reduced to a
Lie subgroup by imposing further constraints on the (algebraic) connection:
trace conditions, projections or even symmetry conditions.\\[36pt]


\end{document}